\documentclass{osa-article}

\journal{oe}

\articletype{Research Article}

\begin{document}

\title{Sensing performance enhancement via asymmetric gain optimization in the atom-light hybrid interferometer}

\author{Zhifei Yu,\authormark{1} Bo Fang,\authormark{1} Pan Liu,\authormark{1} Shuying Chen,\authormark{1} Guzhi Bao,\authormark{2} Chun-hua Yuan,\authormark{1,*} and Liqing Chen  \authormark{1,*}}

\address{\authormark{1}State Key Laboratory of Precision Spectroscopy, Quantum Institute for Light and Atoms, Department of Physics, East China Normal University, Shanghai 200062, China}
\address{\authormark{2}School of Physics and Astronomy, and Tsung-Dao Lee Institute, Shanghai Jiao Tong University, Shanghai 200240, China}
\email{\authormark{*}   chyuan@phy.ecnu.edu.cn} 
\email{\authormark{*}	lqchen@phy.ecnu.edu.cn}


\begin{abstract}
The SU (1,1)-type atom-light hybrid interferometer (SALHI) is a kind of
interferometer that is sensitive to both the optical phase and atomic phase.
However, the loss has been an unavoidable problem in practical applications
and greatly limits the use of interferometers. Visibility is an important
parameter to evaluate the performance of interferometers. Here, we experimentally demonstrate the mitigating effect of the loss on visibility of the SALHI via asymmetric gain optimization, where the maximum threshold of loss to visibility close to $100\%$ is increased. Furthermore, we theoretically find that the optimal condition for the largest visibility is the same as that
for the enhancement of signal-to-noise ratio (SNR) to the best value with
the existence of the losses using the intensity detection, indicating
that visibility can act as an experimental operational criterion for SNR
improvement in practical applications. Improvement of the interference
visibility means achievement of SNR enhancement. Our results provide a significant
foundation for practical application of the SALHI in radar and ranging
measurements.
\end{abstract}

\section{Introduction}
Interferometers are widely used as sensors in precision measurement \cite%
{nie2018experimental,abbott2017gw170817,ligo2017gravitational,liu2017enhancement}%
. There have been many kinds of interferometers, such as optical
interferometers\cite%
{xiao1987precision,bondurant1984squeezed,steuernagel2004approaching}, atom
interferometers\cite%
{gross2010nonlinear,wang2005atom,weitz1994atomic,hamilton2015atoms,estey2015high,kasevich1992measurement}
and atom-light hybrid interferometer (ALHI)\cite{chen2015atom,qiu2016atom}.
Optical interferometers can measure the optical phase sensitive quantitation
and are the core of optical gyroscopes\cite{matsko2004optical}, laser radar%
\cite{rosen2000synthetic}, ranging systems\cite{kubota1987interferometer},
etc. Atom interferometers can measure the atomic phase sensitive parameters,
and have been demonstrated to measure the rotation rate\cite%
{gustavson1997precision}, acceleration of gravity\cite%
{peters2001high,hamilton2015atom,wu2019gravity} and magnetic field\cite%
{zhou2010precisely,ockeloen2013quantum}. The ALHI is sensitive to both the
optical and atomic phases, which has the potential to combine the advantages
of optical wave and atomic systems in precision measurement, and has been
utilized to measure angular velocity, electric field and magnetic field \cite%
{chen2015atom,qiu2016atom,wu2020atom,chen2017quantum}.

In interferometry, the visibility represents the degree of interference
cancellation of two beams, which can evaluate the performance of
interferometers\cite%
{hudelist2014quantum,kong2013experimental,kolobov2017controlling,hamilton2015atom,estey2015high}%
. Low visibility has a negative effect on the measurement and causes a
reduction in the SNR \cite%
{ou2012enhancement,kong2013experimental,kacprowicz2010experimental}. The
SU(1, 1)-type interferometer realizes beam splitting and recombination
through two parametric amplification processes\cite%
{yurke19862,hudelist2014quantum,zheng2020quantum,liu2018loss}, whose two
interference arms are quantum correlated. Comparing with conventional
Mach-Zehnder interferometer (MZI), SNR of SU(1, 1)-type interferometer can
break standard quantum limit due to quantum correlation. Previous
literatures have shown that the SU(1, 1)-type interferometer has the
tolerance to the detection loss (that is, external loss) by increasing the
gain of the wave-recombination process \cite%
{marino2012effect,manceau2017detection,giese2017phase,manceau2017improving,liu2018loss}
. However the SU(1, 1)-type interferometer, the internal loss has a greater
effect on the perfect noise cancellation between two beams\cite%
{xin2016effect,du2020quantum}. Noise cancellation is the advantage of the
SU(1, 1)-type interferometer\cite{li2016phase}. The internal loss limit the
practical application of the SU(1, 1)-type interferometer in radar and
ranging measurements. Recently, in the presence of internal losses, augmenting the visibility through asymmetry is shown in the all optical SU(1, 1) interferometer \cite{Michael21}, and here we extend to the case of SALHI.

In this paper, we experimentally and theoretically investigate the
visibility optimization of SU(1, 1)-type ALHI (SALHI) with the existence of
the internal loss. The conventional MZI is also given as a comparison. The
visibility of SALHI and MZI decreases with the loss. However, we
theoretically give an optimization condition, under which the visibility of
SALHI ($\mathrm{{V_{SU}} }$) can be restored to \symbol{126}100 $\%$ by
optimizing the gain factor of wave recombination process to satisfy the
optimization condition over a wide range of internal loss. In experiment $%
\mathrm{{V_{SU}} }$ is restored to $\sim $90$\%$. Furthermore, we
theoretically analyze the SNR of the SALHI ($\mathrm{{SNR_{SU}} }$) and find
that the optimization condition for $\mathrm{{SNR_{SU}} }$ enhancement is
the same as that for the visibility restoration, which implies that as long
as the $\mathrm{{V_{SU}} }$ is improved, the $\mathrm{{SNR_{SU}} }$ can be
enhanced. In experiment, it is difficult to judge whether the experimental
conditions are suitable to achieve the best $\mathrm{{SNR_{SU}} }$. However,
the visibility is a physical quantity that is convenient to obtain and
observe. Thus, the $\mathrm{{V_{SU}} }$ can be used as an experimental
operational criterion for $\mathrm{{SNR_{SU}} }$ improvement. Therefore, the
optimization condition and the $\mathrm{{V_{SU}} }$ restoration, have
guiding significance for practical application of atom-light hybrid
interferometers in the future.

\label{sec:examples}

\section{SU(1, 1)-type atom-light hybrid interferometer}

\begin{figure*}[t]
	\includegraphics[scale=0.77]{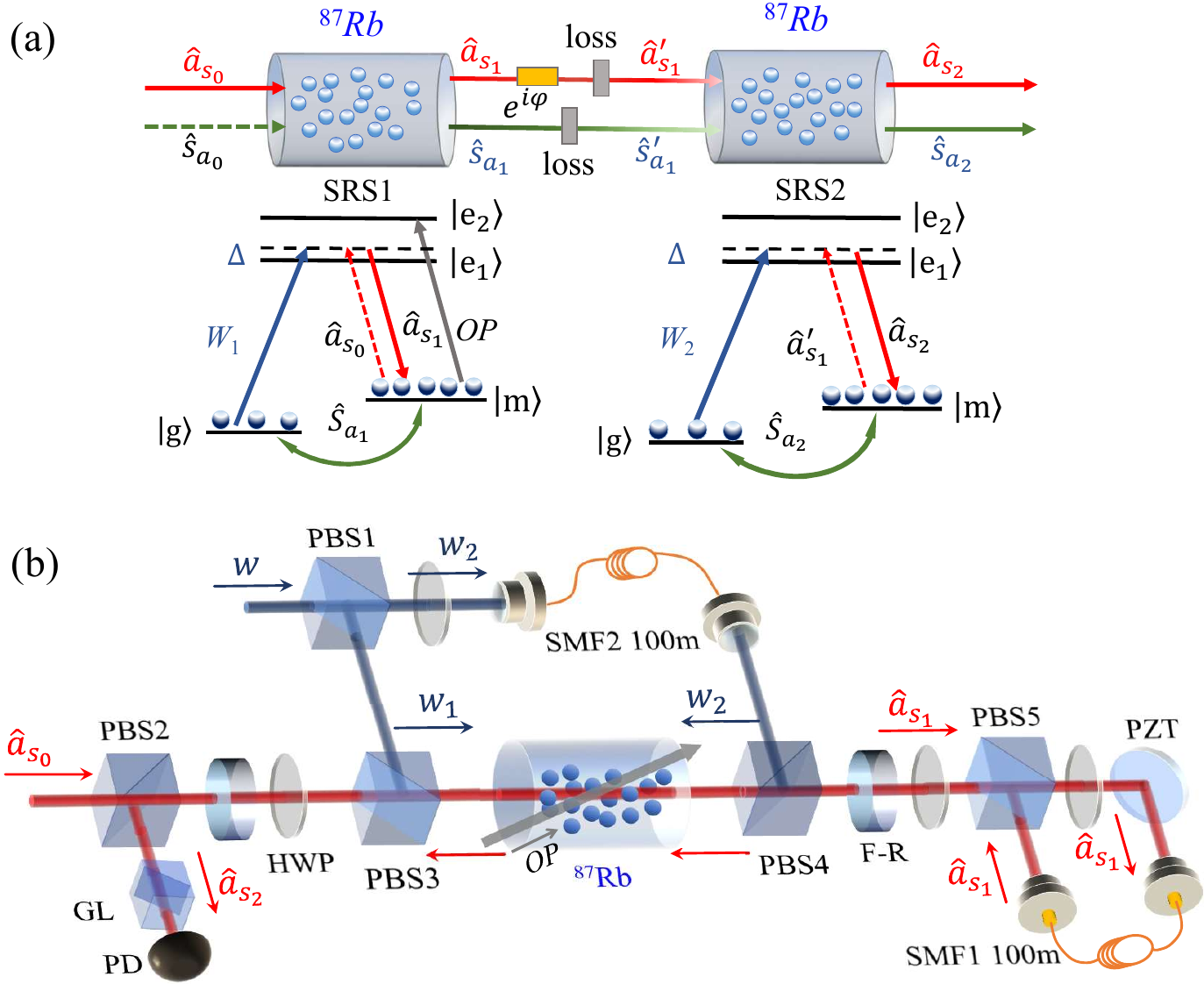}
	\centering
	\caption{(a) Scheme of the SALHI, energy level of the $^{87}Rb$ atom, and
		frequencies of the laser. $\hat{a}_{s_{0}}$ and $\hat{S}_{a_{0}}$ are the
		initially input optical field and atomic spin wave, respectively. $\hat{a}
		_{s_{1}}$ and $\hat{S}_{a_{1}}$ are the interference arms generated from
		SRS1, which then evolve to $\hat{a}^{^{\prime }}_{s_1}$ and $\hat{S}%
		^{^{\prime }}_{a_1}$ after loss and phase shift $\protect\varphi$, finally
		becoming interference outputs $\hat{a}_{s_2}$ and $\hat{S}_{a_2}$ after
		SRS2. SRS: stimulated Raman scattering. $|g,m\rangle $: $%
		|5^{2}S_{1/2},F=1,F=2\rangle $; $|e_{1},e_{2}\rangle $: $|5^{2}P_{1/2},F=2%
		\rangle $, $|5^{2}P_{3/2}\rangle $; $W_1$ and $W_2$: Raman pump fields; OP:
		optical pumping field; $\Delta$: single photon detuning. (b) Experimental
		setup of the SALHI. PZT: piezoelectric transducer; PBS: polarization beam
		splitter; HWP: half-wave plate; F-R: Faraday rotator; GL: Glan prism; SMF:
		single-mode fiber; PD: photon detector. The diameters of $W_1$, $W_2$ and $%
		\hat{ 	a}_{s_{0}}$ are all 0.5mm. }
	\label{fig:epsart}
\end{figure*}

The scheme of the SALHI is shown in Fig. 1 (a), where two stimulated Raman
scattering processes, SRS1 and SRS2, are used to realize the wave splitting
and recombination of optical field and atomic spin wave. SRS1 generates $%
\hat{a}_{s_{1}}$\ and $\hat{S}_{a_{1}}$ acting as two interference arms of
SALHI, and then SRS2 is used to recombine $\hat{a}^{^{\prime }}_{s_{1}}$\
and $\hat{S}^{^{\prime }}_{a_{1}}$. The final interference outputs are
optical signal $\hat{a}_{s_{2}}$ and atomic spin wave $\hat{S}_{a_{2}}$,
respectively.

When SRS is operated in single mode, which can be realized by using the seed
$\hat{a}_{s_0}$ and the $W $ beam in spatial single-mode from single-mode
fiber in experiment. The Hamiltonian of SRS can be written as \cite%
{hammerer2010quantum}
\begin{equation}
\begin{aligned} \hat H=i\hbar\zeta A_W\hat a^\dagger_s\hat S^\dagger_a+H.c,
\end{aligned}  \label{1}
\end{equation}%
where $\zeta =(g_{eg}g_{em})/\Delta $, with $g_{eg}$, $g_{em}$ are the
coupling coefficients and $\Delta $ is the detuning frequency of the $W$
field as Fig. 1 (a) shown. $A_{W}$ is the amplitude of the strong $W$ field.
The input-output relationship of SRS1 is
\begin{equation}
\begin{aligned} \begin{split} \hat{a}_{s_{1}}
&=&G_{1}\hat{a}_{s_{0}}+g_{1}\hat{S}_{a_{0}}^{\dagger }, \\ \hat{S}_{a_{1}}
&=&G_{1}\hat{S}_{a_{0}}+g_{1}\hat{a}_{s_{0}}^{\dagger }, \end{split}
\end{aligned}
\end{equation}%
where $\hat{a}_{s_{0}}$ and $\hat{S}_{a_{0}}$ are the initially input states
of the optical field and atomic spin wave, respectively. $\hat{a}_{s_{0}}$
is the coherent state, and $\hat{S}_{a_{0}}$ is the vacuum state. Between
SRS1 and SRS2, a phase shift $\varphi $, internal loss $l$ of the optical
field, the dephasing $\eta $ of the atomic spin wave are introduced. Then, $%
\hat{a}_{s_{1}}$ becomes $\hat{a}_{s_{1}}^{^{\prime }}=\sqrt{1-l}\hat{a}%
_{s_{1}}e^{i\varphi }+\sqrt{l}\hat{v}$, and $\hat{S}_{a_{1}}$ becomes $\hat{S%
}_{a_{1}}^{^{\prime }}=\sqrt{1-\eta }\hat{S}_{a_{1}}+\sqrt{\eta }\hat{F},$
where $\hat{v}$ and $\hat{F}$ are the operators of vacuum. After the SRS2
process, the interference outputs are
\begin{equation}
\begin{aligned} \begin{split} \hat
a_{s_{2}}=&(G_1G_2\sqrt{1-l}e^{i\varphi}+g_1g_2\sqrt{1-\eta})\hat
a_{s_0}+G_2\sqrt{l}\hat
v\\+&(G_2g_1\sqrt{1-l}e^{i\varphi}+G_1g_2\sqrt{1-\eta})\hat
S_{a_0}^\dagger+g_2\sqrt{\eta}\hat F^\dagger,\\ \hat
S_{a_{2}}=&(G_1g_2\sqrt{1-l}e^{-i\varphi}+G_2g_1\sqrt{1-\eta})\hat
a_{s_0}^\dagger+g_2\sqrt{l}\hat
v^\dagger\\+&(g_1g_2\sqrt{1-l}e^{-i\varphi}+G_1G_2\sqrt{1-\eta})\hat
S_{a_0}+G_2\sqrt{\eta}\hat F, \end{split} \end{aligned}
\end{equation}%
where the Raman gain factors $G_{k}=\frac{1}{2}(e^{\zeta
	_{k}A_{W}}+e^{-\zeta _{k}A_{W}})$ and $g_{k}=\frac{1}{2}(e^{\zeta
	_{k}A_{W}}-e^{-\zeta _{k}A_{W}})$ are related to $A_{W}$ and  $\Delta $ of
the $W$ field. $k=1,2$ represents SRS1 and  SRS2, respectively. $G_{k}$ and $%
g_{k}$ satisfy $G_{k}^{2}-g_{k}^{2}=1$. The outputs $\hat{a}_{s_2}$ and $%
\hat{s}_{a_2}$ both depend on the gain factors, the losses $l$, $\eta$ and
the phase shift.
\section{Experimental setup}

The experiment is performed in a cylindrical paraffin-coated $^{87}\!Rb$
vapor cell (diameter 0.5 cm, length 5 cm). As shown in Fig. 1(b), which was
mounted inside a five-layer magnetic shield to reduce the stray magnetic
field and heated to $75^{\circ }$C. Before the SRS1, almost all atoms are
prepared in the ground state $|g\rangle $ by an optical pumping field (OP)
resonant at the $|m\rangle $ $\rightarrow $ $|e_{2}\rangle $ transition. The
OP pulse is 45 $\mu $s long, and its intensity is 110 mW. The $W$ field is
divided into $W_{1}$ and $W_{2}$. $W_{2}$ is coupled into a 100 m-long
single-mode fiber (SMF2). $W_{1}$ and initial input Stokes seed $\hat{a}%
_{s_{0}}$ are spatially overlapped by PBS3 and interact the atoms via SRS1.
The detuning frequency $\Delta $ of $W$ is 1.2 GHz. The $\hat{a}_{s_{0}}$
beam is red tuned 6.8 GHz from the $W$ laser by an electro-optic modulator
(EOM, Newport model No. 4851). After SRS1, $\hat{S}_{a_{1}}$ stays in the
cell. $W_{1}$ and $\hat{a}_{s_{1}}$ exit the cell and are separated by PBS4.
$\hat{a}_{s_{1}}$ is coupled into 100 m-long SMF1 and then returned back
into the atomic cell with the $W_{2}$ pulse to interfere with $\hat{S}%
_{a_{1}}$ via SRS2. $\hat{a}_{s_{1}}$ and $W_{2}$ are temporally and
spatially overlapped. The phase shift $\varphi $ of $\hat{a}_{s_{1}}$ is
controlled by the PZT. The optical interference output $\hat{a}_{s_{2}}$ is
detected by a photodetector after a Glan prism to filter $W_{2}$. In
experiment, the internal loss $l$ of the optical interference beam is
approximately 0.6, including the coupling efficiency of fiber, the
transmittance of vapor cell and optical devices. The  decay of the atomic
spin wave is 0.4 due to atoms collisions and flying out  of the interacting
region during the evolution time between two SRS  processes.

Comparing to optical MZI, whose two interference arms are both optical waves
and the output is only sensitive to the optical phase, the SALHI goes one
step futher relying atom-photon correlation. Thus the interference fringes
depends on both atomic and optical phases.

\section{Visibility results}

In the experiment, we first measured the $\mathrm{{V_{SU}} }$ and the
visibility of MZI ($\mathrm{{V_{MZ}} }$) with the same phase-sensitive
particle number and internal loss as a comparison. Fig. 2 (a) shows the
interference fringes of the Mach-Zehnder interferometer (MZI) and SALHI at $l
$=0.96, $G_{1}=3$, $G_{2}=5$ and atomic decay rate $\eta =0.4$ of $\hat{S}%
_{a_{1}}$. The values of the $\mathrm{{V_{MZ}} }$ and $\mathrm{{V_{SU}} }$
are 45 $\%$\ and 53 $\%$, respectively. Fig. 2 (b) shows the visibility
value as a function of the loss rate $l$. In general, as the optical loss $l$
of $\hat{a}_{s_{1}}$ increases from 0.6 to 0.96 by variable attenuation
plate, $\mathrm{V_{SU}}$ drops from 92.1$\%$ to 53$\%$, and $\mathrm{V_{MZ}}$
is always smaller than $\mathrm{V_{SU}}$ under the same loss condition.

In theory, according to Eq. (3), the visibility of the optical interference
output $\hat{a}_{s_{2}}$ can be calculated and simplified as
\begin{equation}
\mathrm{V_{SU}}\approx \dfrac{2G_{1}G_{2}g_{1}g_{2}\sqrt{1-l}\sqrt{1-\eta }}{
	G_{1}^{2}G_{2}^{2}(1-l)+g_{1}^{2}g_{2}^{2}(1-\eta )}.
\end{equation}%
$\mathrm{V_{SU}}$ depends not only on the gain factors ($G_{1}$, $g_{1}$, $%
G_{2}$, $g_{2}$) but also on the internal losses $\left( \eta ,l\right) $.
The gain factors can be controlled by SRS parameters, such as the
single-photon detuning $\Delta$ and power of $W$ fields. We give the
theoretical visibility values obtained by using corresponding experimental
parameters ($G_{1}$, $g_{1}$, $G_{2}$, $g_{2}$, $\eta $, $l$) shown in fig.
2 (b) with blue solid lines. The theoretical predictions and experimental
data match well.

\begin{figure}[t]
	\includegraphics[scale=0.64]{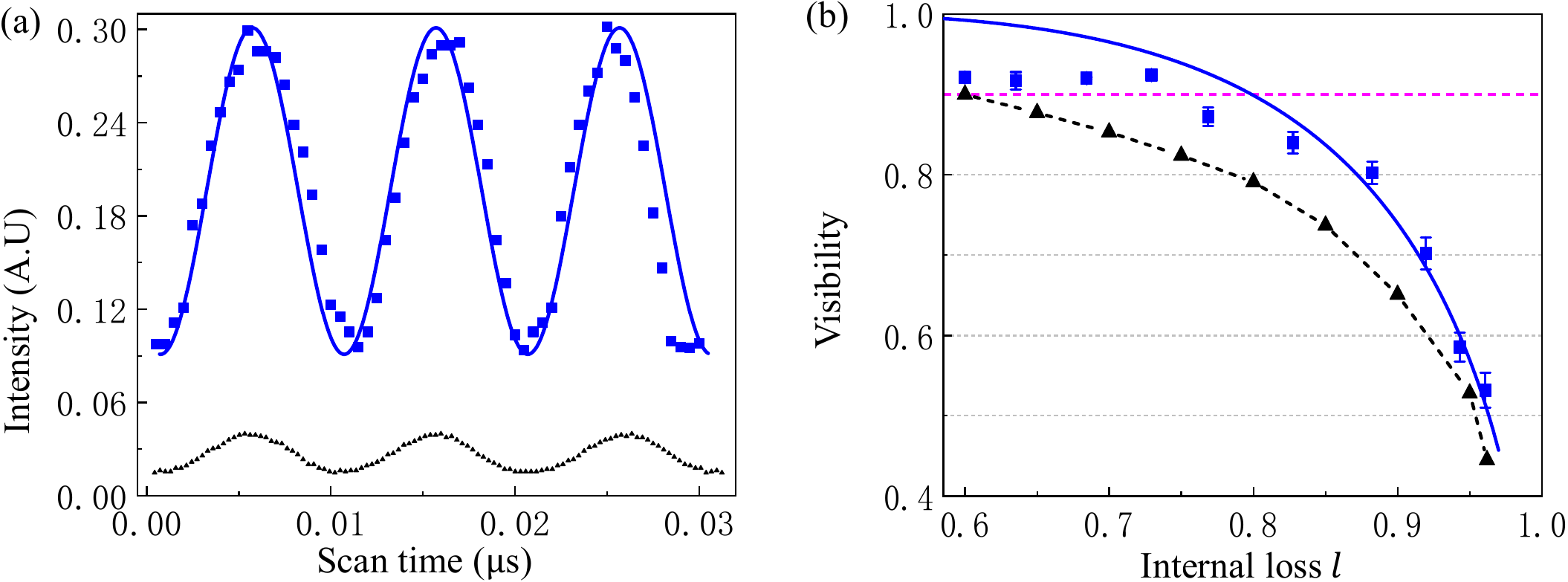}
	\centering
	\caption{ (a) The blue squares are the interference fringes of the SALHI
		with fixed $G_{1}$ =3, $G_{2}$ =5, $l$ =0.96 and $\protect\eta$ =0.4, and
		the black triangles are the interference fringes of the MZI under the same
		operating conditions. (b) The visibility value as a function of the loss
		rate $l$. The blue squares represent $\mathrm{{V_{SU}} }$ with fixed $G_1$ =
		$3 $, $G_2$ = $5$ and $\protect\eta$ = $0.4$, the blue solid curve is the
		result of theoretical fitting according to Eq. (4), and the black triangles
		are the values of $\mathrm{V_{MZ}}$ measured experimentally. The pink dashed
		line corresponds to a visibility of 90$ \% $.}
	\label{fig:epsart}
\end{figure}

\section{Optimization condition}

Furthermore, the largest interference visibility in Eq. (4) appears at
\begin{equation}
G_{1}G_{2}\sqrt{1-l}=g_{1}g_{2}\sqrt{1-\eta }.
\end{equation}%
We call this the optimization condition. According to Eqs. (2, 3), the
interference output $\hat{a}_{s_{2}}$ consists of two parts. One is $G_{2}%
\hat{a}_{s_{1}}^{
	{\acute{}}%
}$ amplified from the optical arm $\hat{a}_{s_{1}}^{ {^{\prime }} }=\sqrt{1-l%
}\hat{a}_{s_{1}}$, and the other is $g_{2}\hat{S}_{a_{1}}^{
	{\acute{}}%
}$ amplified from the atomic arm $\hat{S}_{a_{1}}^{ {^{\prime }} }=\sqrt{%
	1-\eta }\hat{S}_{a_{1}}$. When the gain factor of the wave-recombination
process ($G_2$) is adjusted to satisfy the $G_{1}G_{2}\sqrt{1-l}=g_{1}g_{2}%
\sqrt{ 1-\eta }$, the amplitudes of the two parts are equal, then the
visibility of output $\hat{a}_{s_{2}}$ can reach $\sim $100$\%$.

The SNR is also an important parameter to characterize the performance of an
interferometer and can be calculated by SNR=$\dfrac{[(\partial\langle \hat{O}
	\rangle/\partial\varphi)\delta] ^{2}}{\langle (\Delta \hat{O})^{2}\rangle }$%
\cite{anderson2017phase,chen2016effects}, where $\hat{O}$ is the measurable
operator, $\delta$ is the added modulation small phase and $\langle (\Delta
\hat{O})^{2}\rangle =\langle \hat{O}^{2}\rangle -\langle \hat{O}\rangle ^{2}$%
. Under ID, the $\mathrm{{SNR_{SU}} }$ for the output optical field is

\begin{equation}
\mathrm{SNR_{SU}}\approx \dfrac{%
	4G_{1}^{2}G_{2}^{2}g_{1}^{2}g_{2}^{2}(1-l)(1-\eta )N_{\hat{a}%
		_{s_0}}sin^{2}\varphi \delta ^{2}}{A^{2}(A^{2}+B^{2}+C^{2})},
\end{equation}%
where the input particle number $N_{\hat{a}_{s_0}}$= $\langle \hat{a}%
_{s_{0}}^{+}\hat{a}_{s_{0}}\rangle $, $A^{2}$ = $%
G_{1}^{2}G_{2}^{2}(1-l)+g_{1}^{2}g_{2}^{2}(1-\eta )\\+2G_{1}G_{2}g_{1}g_{2}%
\sqrt{1-l}\sqrt{1-\eta }cos\varphi $, $B^{2}$ = $%
G_{2}^{2}g_{1}^{2}(1-l)+G_{1}^{2}g_{2}^{2}(1-\eta )+2G_{1}G_{2}g_{1}g_{2}%
\sqrt{1-l}\sqrt{1-\eta }cos\varphi $, and $C^{2}$ = $G_{2}^{2}l+g_{2}^{2}%
\eta $. $\mathrm{{SNR_{SU}}}$ is also related to internal losses $\left(
\eta ,l\right) $ and gain factors ($G_{1}$, $g_{1}$, $G_{2}$, $g_{2}$). To
find the best $\mathrm{{SNR_{SU}} }$ condition under a certain loss $l$, we
calculate the partial derivative
\begin{equation}
\dfrac{\partial (\mathrm{{SNR_{SU}})}}{\partial \sqrt{1-l}}=0.
\end{equation}%
When the interferometer operates near the dark point, that is, $\varphi =\pi
+\Delta \varphi $ and $\Delta \varphi \sim 0$, the solution of Eq. (7) is $%
G_{1}G_{2}\sqrt{1-l}=g_{1}g_{2}\sqrt{1-\eta }$, where the best $\mathrm{{%
		SNR_{SU}}}$ can be achieved.

Obviously, this condition for the $\mathrm{{SNR_{SU}} }$ under ID is same as
Eq. (5), indicating that the improvement of $\mathrm{{V_{SU}}}$ corresponds
to enhancement of $\mathrm{{SNR_{SU}}}$. The optimization condition is the
key point to improve $\mathrm{{V_{SU}}}$ and enhance $\mathrm{{SNR_{SU}}}$
even at large internal loss. It should be noted that the interference
visibility can be restored to $\sim 100\%$ and $\mathrm{{SNR_{SU}}}$ can be
enhanced to the best value in the presence of losses when the experimental
conditions satisfy the optimization condition. In the interferometer, phase
shift can be measured using ID and balance homodyne detection (BHD). We also
give the optimization condition for BHD in appendix part, which
is different to Eq. (5).

To show the improvement of the SALHI compared with the conventional MZI
under the same operating conditions, we calculated $\mathrm{{SNR_{MZ}}}$=$%
\dfrac{(1-l)(1-\eta)N_{0}sin^2\varphi\delta^2 }{[(2-l-\eta)-2\sqrt{1-l}\sqrt{%
		1-\eta}cos\varphi]}$\cite{demkowicz2009quantum,dorner2009optimal}, where $%
N_{0}=(2G^2_1-1)N_{\hat{a}_{s_0}}$ is the phase-sensitive particle number of
the MZI. Figs. 3 (a-c) shows the visibility and Figs. 3 (d-f) shows the SNR
as a function of the optical loss $l$. Firstly, before optimization, as the
loss $l$\ increases, the $\mathrm{{SNR_{SU}}}$ first increase to a maximum
value at $l=l_{B}$ and then decrease. In fact, $l_{B}$ is the point
satisfying\textbf{\ }the optimization condition\textbf{\ }$G_{1}G_{2}\sqrt{%
	1-l}=g_{1}g_{2}\sqrt{ 1-\eta }$, and compared with MZI, the quantum
interferometers has better visibility. However, Figs. 3 (d-f) shows that $%
\mathrm{{SNR_{SU}}}$ is larger than $\mathrm{{SNR_{MZ}}}$ only within a
small $l$ range near $l_{B}$ under a certain $G_{1}$, $G_2$ and $\eta $, and
as $G_{1}$ and $\eta $ increase, this range is gradually diminished. The
reason is that the increased $G_{1}$ or internal loss will bring more
uncorrelated excess noise and quickly reduce the noise cancellation
advantage of the SU(1, 1)-type interferometer. Therefore, when $G_{1}$, $%
g_{1}$ and $\eta $ are fixed, finding a suitable $G_{2}$ satisfying the
optimization condition at each $l$ is an effective way to enhance $\mathrm{{%
		SNR_{SU}}>{SNR_{MZ}}}$ over a wider range of internal loss.

\begin{figure*}[t]
	\includegraphics[width=13.2cm,height=7cm]{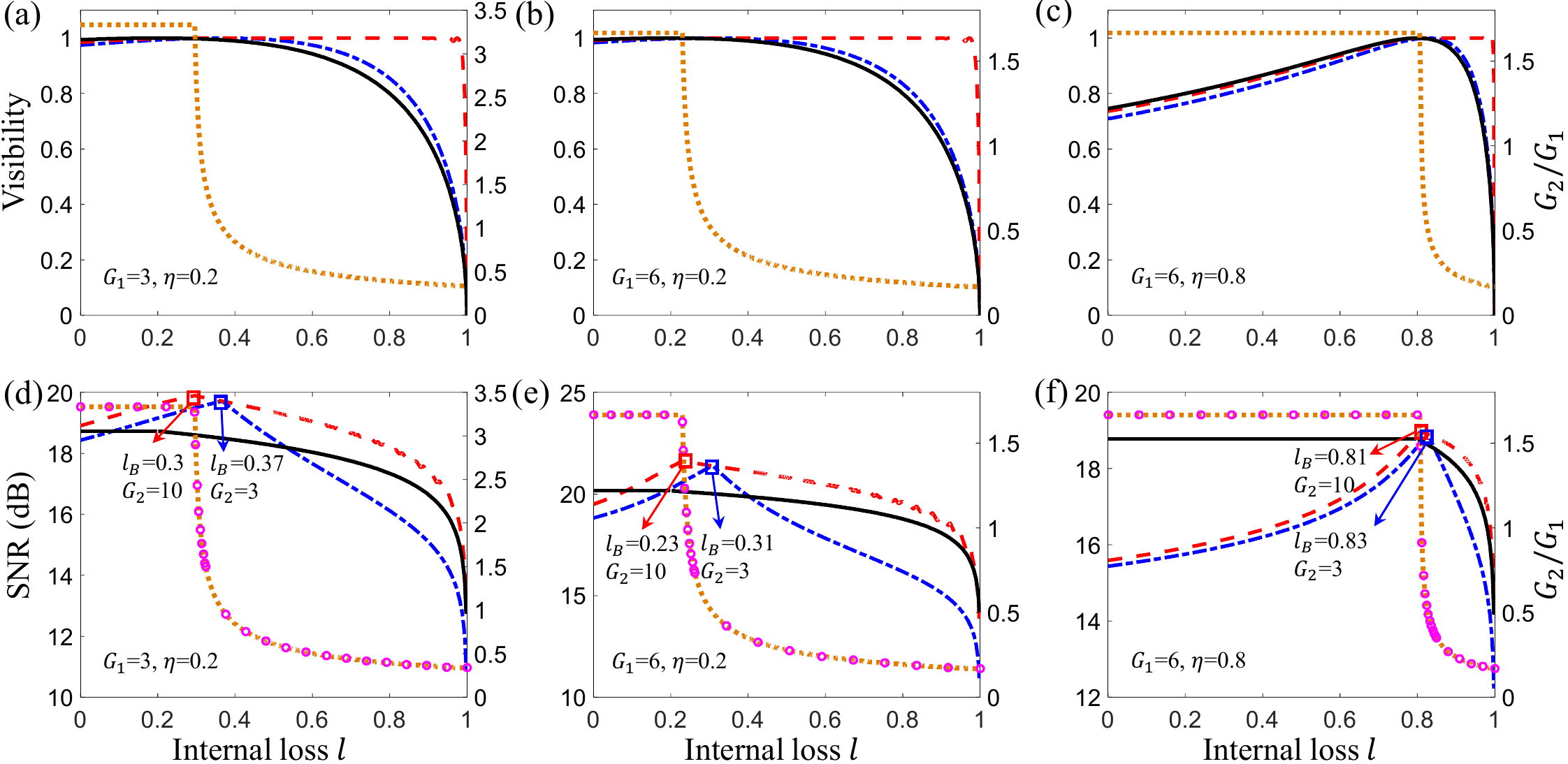}
	\centering
	\caption{ (a-c) The $\mathrm{{V_{SU}}}$ before and after optimization of the
		output field $\hat{a}_{s_2}$ and $\mathrm{{V_{MZ}}}$ as a function of $l$
		(left-hand vertical axis). The blue dash-dotted curve is $\mathrm{{V_{SU}}}$
		before optimization with fixed gain factors $G_{1}$, $g_{1}$, $G_{2}$, and $%
		g_{2}$. The red dashed curve is largest $\mathrm{{V_{SU}} }$ (right-hand
		vertical axis) after
		optimizing $G_2$. The orange dotted curve is the value of $G_{2}/G_{1}$
		after optimizing $G_{2}$ for the largest $\mathrm{{V_{SU}}}$, the black solid line is $\mathrm{{V_{MZ}}}$. (d-f) The black
		solid is $\mathrm{{SNR_{MZ}}}$ and the blue dash-dotted curves is $\mathrm{{%
				SNR_{SU}}}$ before optimization with fixed gain factors $G_{1}$, $g_{1}$, $%
		G_{2}$, and $g_{2}$, respectively, the red dashed curve is $\mathrm{{SNR_{SU}%
		}}$ after optimizing $G_{2}$ (left-hand vertical axis). The pink circles
		represent the $G_{2}/G_{1}$ value at the best $\mathrm{{SNR_{SU}}}$ after
		optimization, and the orange dotted curve is the $G_{2}/G_{1}$ value of the
		largest $\mathrm{{V_{SU}}}$ after optimization (right-hand vertical axis). }
	\label{fig:epsart}
\end{figure*}

Fig. 3 also shows the optimal visibility and $\mathrm{{SNR_{SU}}}$ values
(the left vertical axis) and corresponding $G_{2}$/$G_{1}$ value (the right
vertical axis), $G_{2}$ is limited within 1$\sim $ 10 considering the
experimental operability. First, $\mathrm{{V_{SU}}}$ and $\mathrm{{SNR_{SU}}}
$ after optimization are larger than $\mathrm{{V_{MZ}}}$ and $\mathrm{{%
		SNR_{MZ}}}$ over a wide range of losses. The optimized $G_{2}$ can
effectively reduce the negative impact of the internal loss on $\mathrm{{%
		V_{SU}}}$ and $\mathrm{{SNR_{SU}}}$. Second, the optimized $G_{2}$ value is
different in the regions of $l<l_{B}$ and $l>l_{B}$. For $l>l_{B}$, the
optimal $G_{2}$ value is very small and can be directly calculated according
to fixed $G_{1}$, $g_{1}$, $\eta $, and $l$. For $l<l_{B}$, we can not
obtain the $G_{2}$ value completely satisfying the condition, only a larger $%
G_{2}$ value is closer to satisfying. Therefore as $l$ increases, there is a
common feature in Figs. 3 (a-c) that the optimized $G_{2}$/$G_{1}$ value
first remains at the maximum value at $\leq l_{B}$ and then decrease sharply
to a much smaller value at $l\geq l_{B}$. Finally, in Figs. 3 (d-f), the
pink circles and the orange dotted curve completely coincide (the right
vertical axis), showing that at any internal loss, the best $\mathrm{{%
		SNR_{SU}}}$ corresponds to the point of largest $\mathrm{{V_{SU}}}$.
Optimization of $\mathrm{{V_{SU}}}$ is easy to observe in an experiment. As
long as the maximum of $\mathrm{{V_{SU}}}$ is observed by optimizing $G_{2}$%
, we can guarantee the best performance of the SU(1, 1)-type interferometer.
This is different from the approach to compensate for the impact of the
external loss.

\begin{figure}[t]
	\includegraphics[scale=0.65]{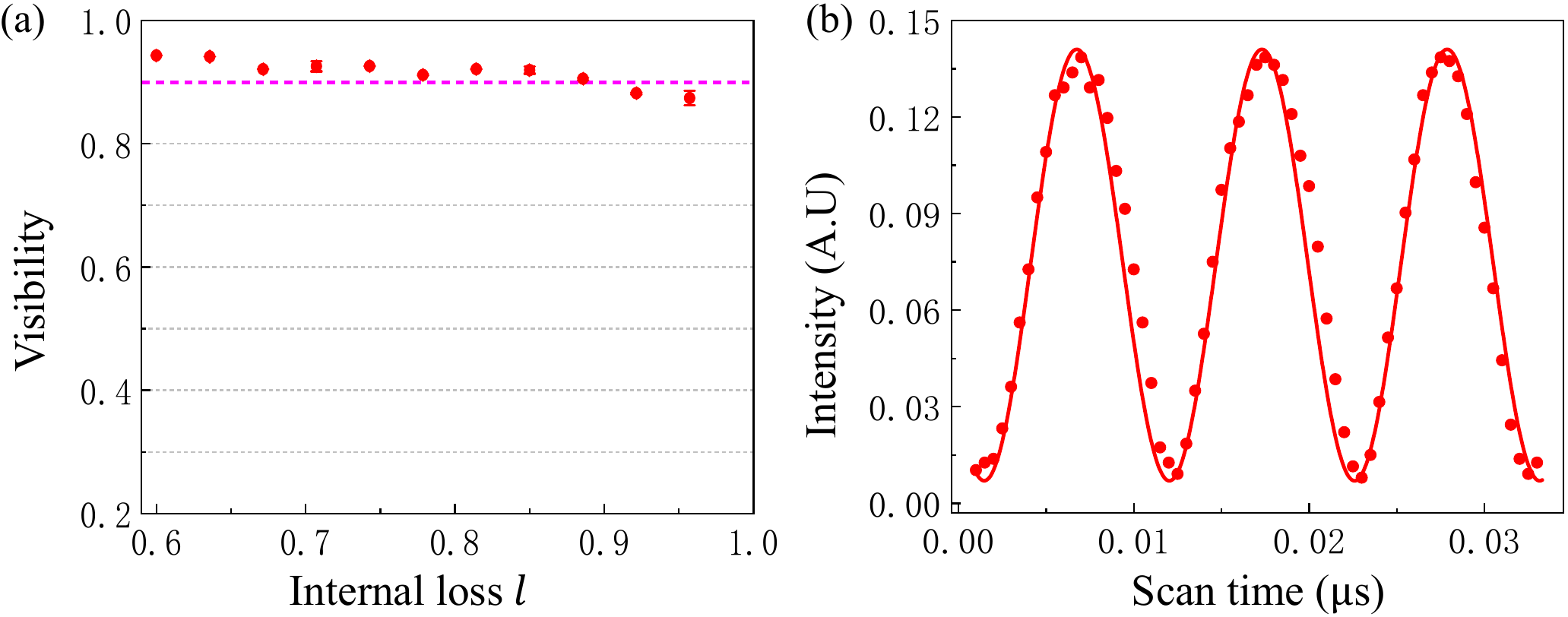}
	\centering
	\caption{(a) The red dots are the values of $\mathrm{{V_{opt}}}$. The pink dashed line corresponds to a visibility of $90\%$. (b) The
		red dots are the interference fringes of the SALHI after optimization,
		corresponding to $G_2$$\approx$1.3 with $G_1$=3, $l$=0.96 ,and $\protect\eta$%
		=0.4, $\mathrm{{V_{opt}}=88\%}$. }
	\label{fig:epsart}
\end{figure}

\section{Visibility restoration}

Next, we provide an experimental demonstration of restoration of $\mathrm{{%
		V_{SU}}}$ by optimizing $G_{2}$ in the SALHI. In practical applications such
as radar or ranging measurement, when the parameters ($G_{1}$, $g_{1}$, $%
\eta $, $l$) are fixed, we can adjust only $G_{2}$ and $g_{2}$ to satisfy
the optimization condition and improve the visibility. In the experiment, $%
W_{1}$ and $W_{2}$ are separated from the same laser as Fig. 1 (b) shown.
Before the $W_{2}$ field enters the vapor cell, it passes through an
attenuator and an AOM, which can be used to adjust the intensity and
frequency of the $W_{2}$ field, respectively. Therefore, we can control $%
G_{2}$ by controlling the intensity and frequency of $W_{2}$, so that the
optimal condition is satisfied to obtain the best visibility. The
experimental data of the visibility after optimizing $G_{2}$ ($\mathrm{{%
		V_{opt}}}$) are given in Fig. 4 (a) using red dots.
$\mathrm{{V_{opt}}}$ is larger than the visibility without optimization ($%
\mathrm{{V_{SU}}}$), and at $l=0.6$, $\mathrm{V_{opt}}$ is 95\%. As $l$
increases, $\mathrm{{V_{opt}}}$ can remain at $\sim 90\%$ (see the pink dashed line), and the optimized
$G_{2}$ is small at the loss $l$ of 0.6-0.96 as in the theoretical
prediction, such as $\mathrm{{V_{opt}}}$=88\% with $G_{2}\approx 1.3$ when $%
l=0.96$, and Fig. 4 (b) shows the interference fringes of the SALHI after
optimization. The results show that the visibility can be restored even at
large internal loss by optimizing $G_{2}$, so the negative impact of
internal loss on the properties of the SALHI can be mitigated. These
experimental results are well consistent with the theoretical expectations.

\section{Discussion and Conclusion}

In conclusion, we have experimentally and theoretically researched the
influence of the internal loss on the visibility of the SU(1, 1)-type
ALHI. In general, the internal loss has a
significant negative impact on the visibility. Moreover, we give the
optimization condition $G_{1}G_{2}\sqrt{1-l}=g_{1}g_{2}\sqrt{1-\eta }$ for
visibility restoration and experimentally demonstrate that the visibility
can be restored to $\sim $90$\%$ over a large range of internal loss by
optimizing the $G_{2}$ factor to satisfy the optimization condition.
Finally, we also theoretically find that the optimization condition for $%
\mathrm{{SNR_{SU}}}$ enhancement is the same as that for visibility
restoration. Visibility as a physical quantity that is easy to obtain and
observe, which can be used as an experimental operational criterion to judge
whether the $\mathrm{{SNR_{SU}}}$ is optimized. What we have found will guide
significance for practical application of quantum measurement.

\section{APPENDIX: Comparison of optimization conditions of ID and BHD}

\textbf{1.  The optimization conditions of seed light field $\hat{a}_{s_0}$ input}

After considering the losses $ l $ and $ \eta $, the optimal condition for the best $\rm {SNR_{SU}} $ under the ID is $  G_{1}G_{2}\sqrt{1-l}=g_{1}g_{2}\sqrt{1-\eta} $,  Obviously, this corresponds to the condition of the largest visibility. However, under the BHD, the quadrature component of interference output at  phase dark point $ \varphi=0 $ is $\hat{X}_{\hat
	a_{s_{2}}}$, 
\begin{equation}
\begin{aligned} \begin{split} 
\hat{X}_{\hat
	a_{s_{2}}}=&(G_1G_2\sqrt{1-l}+g_1g_2\sqrt{1-\eta})\hat
a_{S_0}+G_2\sqrt{l}\hat
v\\+&(G_2g_1\sqrt{1-l}+G_1g_2\sqrt{1-\eta})\hat
S_{a_0}^\dagger+g_2\sqrt{\eta}\hat F^\dagger\\+& 
(G_1G_2\sqrt{1-l}+g_1g_2\sqrt{1-\eta})\hat
a_{S_0}^\dagger+G_2\sqrt{l}\hat
v^\dagger\\+&(G_2g_1\sqrt{1-l}+G_1g_2\sqrt{1-\eta})\hat
S_{a_0}+g_2\sqrt{\eta}\hat F,
\end{split} \end{aligned}
\end{equation}
therefore, under the BHD,
\begin{equation}
\begin{aligned} \begin{split}
{\rm{SNR_{SU}}}=\dfrac{4(1-l)G^{2}_1G^{2}_2N_{\hat{a}_{s_0}}\delta^2}{\zeta^2_1+\zeta^2_2+\zeta^2_3},
\end{split} \end{aligned}
\end{equation}
where $ \zeta^2_1 $=$ (G_1G_2\sqrt{1-l}+g_1g_2\sqrt{1-\eta})^2 $, $ \zeta^2_2 $=$ (G_2g_1\sqrt{1-l}+G_1g_2\sqrt{1-\eta})^2 $,  $ \zeta^2_3 $=$ G^2_2l+g^2_2\eta $.
We also calculate the partial derivative to find the optimization condition, the result is,
\begin{equation}
\begin{aligned} \begin{split}
2\sqrt{1-l}\sqrt{1-\eta}G_1G_2g_1g_2=2(1-\eta)g^2_1g^2_2+g^2_2+G^2_2.
\end{split} \end{aligned}
\end{equation}
Obviously, this is different with the optimization  condition of the largest visibility in Eq. (5).
\\

\textbf{2. The optimization conditions of  initially prepared spin wave $\hat{S}_{a_0}$}

From the Eq.(3), the visibility expression is:
\begin{equation}
{\rm V_{SU}}\approx \dfrac{2G_{1}G_{2}g_{1}g_{2}\sqrt{1-l}\sqrt{1-\eta }}{%
	G_{2}^{2}g_{1}^{2}(1-l)+G_{1}^{2}g_{2}^{2}(1-\eta )}.
\end{equation}
Similarly, for largest visibility, the optimization condition is,
\begin{equation}
G_{2}g_{1}\sqrt{1-l}=G_{1}g_{2}\sqrt{1-\eta}.
\end{equation}  
Under the ID,
\begin{equation}
{\rm SNR_{SU}}\approx \dfrac{4G_{1}^{2}G_{2}^{2}g_{1}^{2}g_{2}^{2}(1-l)(1-\eta
	)N_{\hat{a}_{s_0}}sin^{2}\varphi \delta ^{2}}{B^{2}(A^{2}+B^{2}+C^{2})},
\end{equation}
from the Eq.(7) we can get the optimization condition of best $ {\rm SNR_{SU}} $ is $ G_{2}g_{1}\sqrt{1-l}=G_{1}g_{2}\sqrt{1-\eta} $, which is also same as optimization of largest visibility in Eq. (12).

However, under the BHD, 
\begin{equation}
\begin{aligned} \begin{split}
{\rm{SNR_{SU}}}=\dfrac{4(1-l)G^{2}_2g^{2}_1N_{\hat{a}_{s_0}}\delta^2}{\zeta^2_1+\zeta^2_2+\zeta^2_3},
\end{split} \end{aligned}
\end{equation}
the optimization condition of best $ {\rm SNR_{SU}} $ is same as Eq. (10), which is also different with the optimization condition of largest visibility in Eq. (12).

In previous paper\cite{chen2015atom}, we theoretically studied the $ {\rm{SNR_{SU}}} $ using homodyne detection only considering optical loss $ l $. In this paper, we further study the visibility and $ {\rm{SNR_{SU}}} $ using ID and BHD with both losses $ l $ and $\eta$ because these two losses are always exist simultaneously in practical application. We find that whether with optical input seed or initial atomic seed, the optimization condition for best $ {\rm{SNR_{SU}}} $ using ID is same as that of largest visibility, but different with that using BHD. 

Therefore, here we experimentally measure the signal using ID. We can intuitively judge whether the optimization conditions for best $ {\rm{SNR_{SU}}} $ is achieved  according to the visibility restoration. And furthermore, compared with BHD, the ID device is simpler and more suitable for practical application of the SALHI in radar and ranging measurements.

\begin{backmatter}
\bmsection{Funding}
This work was supported by the National Key Research and Development Program
of China (2016YFA0302001); the National Natural Science Foundation of China
(11874152, 11974111, 11654005, 91536114); the Shanghai Municipal Science and
Technology Major Project (2019SHZDZX01); the innovation Program of Shanghai
Municipal Education Commission (No. 202101070008E00099); the Fundamental
Research Funds for the Central Universities; the Shanghai talent program and
the Fellowship of China Postdoctoral Science Foundation (2020TQ0193).

\bmsection{Disclosures}
The authors declare no conflicts of interest.

\bmsection{Data Availability }
Data underlying the results presented in this paper are not publicly available at this time but may
be obtained from the authors upon reasonable request.

\end{backmatter}


\bibliography{ref}






\end{document}